\documentclass[prl,twocolumn,showpacs,nobibnotes,floatfix,superscriptaddress]{revtex4}

\usepackage{graphicx,eucal}

\begin{document}


\title{Quantum Cellular Automata Pseudo-Random Maps}

\author{Yaakov S. Weinstein}
\thanks{To whom correspondence should be addressed}
\email{weinstei@dave.nrl.navy.mil}
\author{C. Stephen Hellberg}
\email{hellberg@dave.nrl.navy.mil}
\affiliation{Center for Computational Materials Science, Naval Research Laboratory, Washington, DC 20375 \bigskip}

\begin{abstract}
Quantum computation based on quantum cellular automata (QCA) can greatly 
reduce the control and precision necessary for experimental implementations
of quantum information processing. A QCA system consists of a few species 
of qubits in which all qubits of a species evolve in parallel. We show that, 
in spite of its inherent constraints, a QCA system
can be used to study complex quantum dynamics. 
To this aim, we demonstrate scalable operations on a QCA system that fulfill 
statistical criteria of randomness and explore which criteria of randomness 
can be fulfilled by operators from various QCA architectures. 
Other means of realizing random operators with only a few independent operators
are also discussed. 
\end{abstract}

   
\maketitle

The traditional approach to quantum computation has been through the 
circuit model: a series of one and two-qubit gates are applied to 
specified qubits in a specified order \cite{D}. Such an architecture 
requires exquisite control of each individual qubit and accurate localization 
of the external Hamiltonian. An alternative approach to quantum computation 
utilizes quantum cellular automata (QCA). A QCA system consists of just a few
(typically 1-3) species of qubits such that all qubits of a species are 
addressed equivalently and simultaneously. 

The idea of using a QCA system for quantum computation was introduced by
Lloyd \cite{Seth} more than a decade ago. Lloyd demonstrated the universality 
of a three-species QCA and provided pulse sequences for fundamental gates. 
Further work on QCA has concentrated on proving universality \cite{W},
including the universality of a two species QCA that is unable 
to distinguish the left neighbor from the right \cite{Benj}.
Only recently has there been an attempt to exploit the uniqueness of the 
QCA architecture to enhance quantum information processing 
protocols. Brennen and Williams \cite{Bren} demonstrated the production and 
manipulation of entanglement in a QCA architecture. In this work we attempt to 
utilize the QCA architecture in the study of complex quantum dynamics.    

Classical cellular automata (CCA) are systems that follow a simple set of 
local rules applied uniformly to a lattice of cells \cite{Wolf}. Each cell 
can have an arbitrary number of possible states and, at every time interval, 
the state of each cell is updated based on its current state and the state 
of a given radius of nearest neighbors. For example, the evolution of a 
two-state, radius one, one-dimensional CCA updates cell $j$ based on its own 
state and the state of its nearest neighbors. In this case, the evolution has 
eight update directives, one per combination of the three two-state cells, 
$j-1$, $j$, and $j+1$. All cells in a CCA evolve in parallel. This is done 
by copying the current CCA state for the cells to refer to when updating. 
Though an apparently simple system, CCA rules can develop complex 
behavior and can simulate a wide range of phenomena from lattice gases 
to traffic flow.  

A QCA consists of a lattice of 
quantum cells, each with an arbitrary number of levels. The dynamics 
of each cell can depend on a given radius of nearest neighbors, but is 
restricted by the requirement of unitary dynamics. In addition, the 
no-cloning rule outlaws parallel updating. The latter obstacle can be 
overcome by using at least two species of qubits and updating each 
species separately \cite{Bren}. In this work we assume two-level quantum 
cells, referred to as qubits. In addition, we explore 
only radius one evolution, in which each qubit interacts only with its nearest 
neighbors.

Given that CCA are valuable in the simulation of complex classical 
systems, it is natural to ask whether a QCA could be used to simulate complex
quantum systems. Of course, a QCA
that is universal can simulate any dynamics. The question is whether 
the unique architecture of the QCA can provide a less taxing experimental 
venue or provide further insight into complex dynamics. As a first step 
towards answering these questions, we explore the ability of a QCA to 
implement random unitary operators efficiently.

Random matrices were introduced by Wigner to describe the energy levels
of atomic nuclei \cite{Wigner}. Since then, random matrices have been used as
statistical models for a host of complex systems in many areas of physics 
\cite{RMT}. From a quantum computation standpoint, some of the important 
systems modeled via random matrices include quantum systems whose classical 
dynamics are chaotic\cite{QC}, decoherence \cite{Gorin}, and quantum computer 
error models \cite{QCE}. Thus, the ability to implement a random unitary 
operator allows for the simulation and study of these types of systems. 

Beyond simulations, random unitary operators and random quantum states,
created by applying a random unitary to a computational 
basis state, play a vital role in quantum communication and computation. 
In quantum communication, random states are known to saturate the
classical communication capacity of a noisy quantum channel \cite{Seth2}. 
In addition, superdense coding of quantum states \cite{Aram}, a reduction 
in key length for approximate encryption of quantum states, and the 
construction of more efficient data hiding schemes are protocols 
enabled by random unitary operators \cite{Hayden}. Random unitaries 
can also decrease the classical communication cost for remote state 
preparation \cite{Bennet}. Quantum computing protocols facilitated by 
random unitaries include quantum process tomography via a fidelity
decay experiment using a random operator. Since random unitaries will not
commute with noise sources effecting the system, they can identify
the type and strength of the noise \cite{RM}. Random quantum states can be 
used for unbiased sampling, and the amount of multi-partite entanglement
in random states approaches the maximum at a rate exponential with
the number of qubits in the system \cite{Scott}.

The appropriate measure against which random unitary operators and quantum
states are defined is the Haar measure on the group $U(N)$, where $N$ is the 
dimension of Hilbert space. The random ensemble of unitaries drawn from 
this measure is the circular unitary ensemble (CUE) \cite{Mehta}. 
Unfortunately, an exact parameterization of CUE via the Hurwitz
decomposition \cite{Zyc} requires exponential computing resources. 
Recently, however, pseudo-random unitary operators were introduced as 
efficiently implementable substitutes that fulfill various criteria of 
randomness and can be used in the above mentioned protocols \cite{RM}. 

The algorithm to produce pseudo-random operators, or maps, consists of $m$ 
iterations of the $n$ qubit gate: apply a random SU(2) rotation 
on each qubit, then evolve the system via all nearest neighbor 
couplings \cite{RM}. A random SU(2) rotation on qubit $j$ during 
iteration $i$ is defined as 
\begin{eqnarray}
R(\theta^j_i,\phi^j_i, \psi^j_i) &=&
\left(
\begin{array}{cc}
e^{i\phi^j_i}\cos\theta^j_i & e^{i\psi^j_i}\sin\theta^j_i \\ 
-e^{-i\psi^j_i}\sin\theta^j_i &  e^{-i\phi^j_i}\cos\theta^j_i \\
\end{array}
\right), 
\end{eqnarray}
and the nearest neighbor coupling operator is
\begin{equation}
\label{nnc}
U_{nnc} = e^{i(\pi/4)\sum^{n-1}_{j=1}\sigma_z^j\otimes\sigma_z^{j+1}},
\end{equation}
where $\sigma_z^j$ is the $z$-direction Pauli spin operator.
The random rotations are different for each qubit and each iteration, but the 
coupling is always $\pi/4$ to maximize entanglement. After the $m$ 
iterations, a final set of random rotations is applied. 

This paper suggests a modified version of the above algorithm that can 
generate pseudo-random maps applicable to a QCA. The ability to efficiently
generate such operators indicates that complex systems 
can be modeled and explored on a QCA. Moreover, the number 
of iterations needed to create the pseudo-random operators for QCA is 
comparable the number needed for algorithms using a circuit model 
architecture. 

The modification of the algorithm for application to a QCA system is 
straightforward. For each iteration, apply species specific random 
rotations followed by nearest neighbor coupling. Thus, iteration $i$ 
of a QCA random map consists of applying random rotation $U^A_i$ on 
all qubits of species $A$, followed by a different random rotation, 
$U^B_i$, applied to all qubits of species $B$, and so on for all $k$ 
species of qubits, followed by $U_{nnc}$, given in equation \ref{nnc}. 
In keeping with the original 
algorithm a final rotation of the qubits is always applied.

This work may be viewed from a different perspective: 
an examination of how difficult (or easy) it is to create 
pseudo-random operators. The algorithm of \cite{RM} requires 
$(3mn+1)$ independent variables for a pseudo-random operator.
Three independent variables per iteration for each random 
rotation, and one more for the coupling constants. However, as noted in
\cite{RM}, any universal gate set, no matter how biased, 
asymptotically generates the uniform measure of unitary operators. 
This does not imply that a universal gate set will efficiently generate 
the uniform measure of unitary operators. Nor does it suggest that a 
non-universal gate set cannot display some characteristics of randomness. 
Here, we attempt to reduce the number of independent variables required 
for the pseudo-random operator algorithm and see if it is still possible 
to efficiently generate CUE-like statistics. If not, we explore whether 
the generated operators demonstrate any characteristics of randomness. 
The modified pseudo-random operator algorithm for QCA requires only 
$3mk+1$ independent variables. Other possible ways of reducing the number 
of independent variables will also be discussed. 

It is important to state that reducing the number of independent 
variable does not necessarily reduce the experimental difficulty 
in realizing the algorithm. Rather, experimental realizations would 
tend not to exhibit the symmetries that arise by reducing the number 
of independent variables as this would require acute precision over 
the internal and external system Hamiltonians. In this work, the 
first attempts at randomness are always via operators with maximum 
symmetries as these are the most difficult cases with which to achieve 
randomness. 

Throughout this paper we avoid specifying the actual quantum computing
hardware beyond the assertion of a $\sigma_z^j\sigma_z^{j+1}$
coupling between nearest neighbor qubits. The $\sigma_z^j\sigma_z^{j+1}$
interaction is used in the original pseudo-random
operator algorithm and is appropriate for certain proposed realizations 
of quantum information processing. Systems such as quantum dots,
however, interact via the Heisenberg interaction, 
${\bf\hat{s}^j} \cdot {\bf\hat{s}^{j+1}}$, which has a total
spin symmetry. Rotating all the qubits in parallel, as done for the 
$k = 1$ QCA, commutes with the Heisenberg interaction 
and no randomness will be generated. This symmetry can be broken and 
random statistics regained  with a two-species QCA. Moreover, 
the spin-orbit coupling present in actual quantum dots, generally 
considered a negative and worthy of cancellation\cite{so}, introduces 
anisotropy into the system, breaking the symmetry and allowing for 
randomness even with the $k=1$ QCA.  

The introduction of different random rotations at each 
iteration is not in accordance with typical CCA evolution, which follows the 
same rule at each time step. Nonetheless, the advantages of the QCA 
architecture, namely the reduced need of external Hamiltonian localization 
and control requirements, are still manifest in this algorithm.

The first QCA we explore is a single $k=1$ species chain. Every qubit 
in the chain rotates in parallel at every iteration. This map requires 
only $3m+1$ independent 
variables and has an inherent mirror symmetry stemming from the 
equivalence of evolution for all qubits barring those at the edges 
of the chain. Perhaps surprising is that, despite the simplicity 
of the system and the inherent mirror symmetry of the operators, 
the operators fulfill several statistical measures of randomness. 

\begin{figure}
\includegraphics[height=5.8cm, width=8cm]{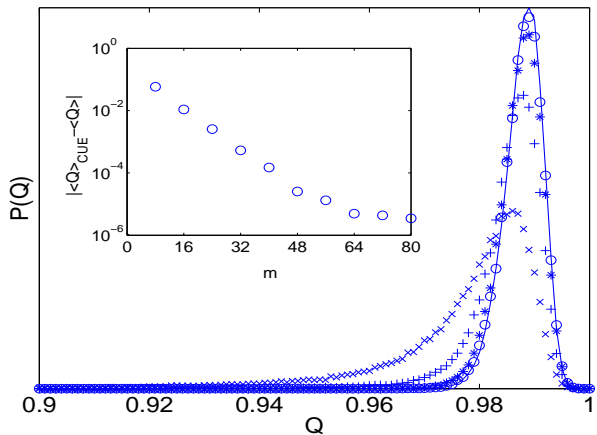}
\caption{\label{QQCA} 
Distribution of the multi-partite entanglement, $Q$, for $m = 16$ (x), 24, (+),
32 (*), and 40 (o) compared to the distribution of $Q$ for CUE random 
operators (solid line). $Q$ is calculated for the wavefunctions produced 
from evolving 
all (256) possible computational basis states under each of 500 eight-qubit 
maps for each $m$ value. For $m = 40$ the distribution is barely 
distinguishable from the CUE distribution. These distributions are 
similar to those found in \protect\cite{RM} despite the further 
constraints imposed by the QCA architecture. The inset shows the 
difference between the average $Q$ of states from the QCA 
operators and operators drawn from CUE as a function of $m$. 
}
\end{figure}

As mentioned above, the production of entanglement is one of the motivations 
for implementing random unitary operations. 
Figure (\ref{QQCA}) shows the distribution of $Q$, the multi-partite 
entanglement measure \cite{Meyer,Bren2}, for states produced from QCA 
maps operating on computational basis states: 
\begin{equation}
Q = 2-\frac{2}{n}\sum^n_{j=1}Tr[\rho_j^2],
\end{equation}
where $\rho_j$ is the reduced density matrix of qubit $j$. 
As $m$, the number of iterations, increases, the distribution approaches 
that of CUE operators acting on the same states. For $m = 40$ the distribution 
of $Q$ is practically indistinguishable from that of CUE \cite{Scott}.
This is the same value of $m$ necessary for the circuit model pseudo-random
operator algorithm to produce a similar distribution.

\begin{figure}
\includegraphics[height=5.8cm, width=8cm]{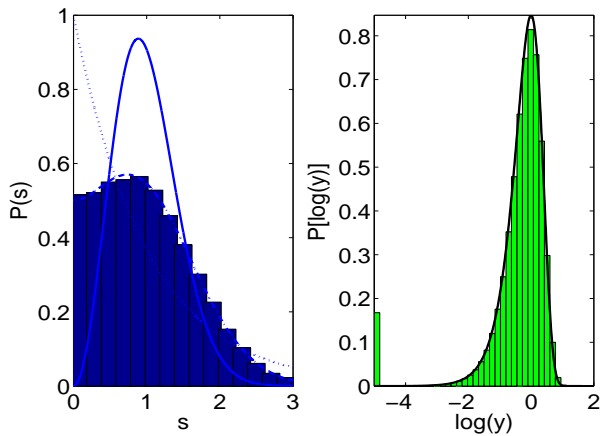}
\caption{\label{MQCA} 
Statistical measures of randomness for 100 one-species, $m = 40$,
eight-qubit QCA maps. The left figure shows the distribution of $s$, the 
nearest neighbor eigenangle spacings compared to that expected for 
random unitary operators (solid line), regular integrable operators 
(dotted line), and a CUE operator with mirror symmetry (dash-dot line). 
The distribution follows that expected for a random operator with 
mirror symmetry. The right figure shows the distribution of the 
magnitude of the eigenvector elements $y$ compared to the distribution 
expected from CUE. There is a noticeable deviation from the random 
distribution. This is seen most clearly by the large number of very small 
terms, $y < 10^{-5}$, depicted by the spike at the left of the figure.  
Nevertheless, the fidelity decay behavior of the maps 
(not shown), which depends on the eigenvector statistics, follows the 
expected exponential of random maps.
}
\end{figure}

Though the maps generated for the QCA, $k = 1$ architecture follow the CUE
entanglement distribution, they deviate from the 
expected random statistics for other important distributions. 
Perhaps the most widely used statistic for the randomness of operators
are the spacings between nearest neighbor eigenvalues (or in the case of 
unitary matrices, eigenangles). For matrices of the CUE the expected 
distribution is \cite{Mehta}:
\begin{equation}
P_{CUE}(s) = \frac{32s^2}{\pi^2}e^{4s^2/\pi}.
\end{equation}
where $s$ is the difference between two ordered eigenvalues. The mirror 
symmetry of the system insures that the operator eigenvalues will not 
follow this distribution. Rather, the statistics follow the superposition 
of two independent CUE spectra, as shown in figure (\ref{MQCA}). 
The total distribution for a matrix of two unequal size blocks both with 
CUE distribution, derived from \cite{spec2}, is
\begin{eqnarray}
P_{CUE}^{(2)}(s,g_1,g_2) & = & 2g_1g_2[1-{\rm erf}_1-{\rm erf}_2+{\rm
erf}_1{\rm erf}_2]\nonumber\\
&+&\frac{32}{\pi^2}s^2e^{-4(g_1^2+g_2^2)s^2/\pi}(g_1^4+g_1^2g_2^2+g_2^4)
\nonumber\\
&+& \frac{8}{\pi}g_1g_2s\Big[g_1e^{-4g_1^2s^2/\pi}(1-g_1^2\frac{4s^2}{\pi}){\rm
erfc}_2 \nonumber\\
&+& g_2e^{-4g_2^2s^2/\pi}(1-g_2^2\frac{4s^2}{\pi}){\rm erfc}_1\Big],
\end{eqnarray}
where $g_i$ is the fraction of Hilbert space spanned by block $i = 1,2$,
and ${\rm erf}_i = {\rm erf}(\frac{2g_is}{\sqrt{\pi}})$ and 
${\rm erfc}_i = {\rm erfc}(\frac{2g_is}{\sqrt{\pi}})$ are the error 
function and complementary error function respectively.
For an 8 qubit operator with mirror symmetry, $g_1 = 15/32$, $g_2 = 17/32$,
and the resulting distribution is barely distinguishable from the $g_1 = g_2 = 1/2$ case,
\begin{eqnarray}
P_{CUE}^{(2)}(s) &=& 
\frac{1}{2}{\rm erfc}(\frac{s}{\sqrt{\pi}})^2 + \frac{6}{\pi^2}s^2e^{-2s^2/\pi}\nonumber\\
&+& \frac{2}{\pi}se^{-s^2/\pi}(1-\frac{s^2}{\pi}){\rm erfc}\left(\frac{s}{\sqrt{\pi}}\right).
\end{eqnarray}

The elements of the eigenvectors of random operators also follow ensemble 
specific distributions. For CUE the appropriate distribution, as 
$N\rightarrow\infty$, is \cite{Mehta}:
\begin{equation}
P_{CUE}(y) = e^{-y}
\end{equation}
where $y = N\eta$, and $\eta$ is the squared modulus of the eigenvector 
element. The randomness of the eigenvector elements determines the 
systems response to perturbation in the sense of fidelity decay \cite{J}. 
The eigenvector element distribution of $k = 1$ QCA maps deviate 
slightly from the CUE distribution, as seen in figure (\ref{MQCA}). 
Nevertheless, these maps demonstrate the exponential decay of fidelity 
at the rate expected for CUE operators.    

To summarize, while these operators are not 
random with respect to the Haar measure, they appear sufficiently
random for entanglement production. They may also be used for protocols 
relying on randomness of eigenvectors such as random operator quantum 
process tomography \cite{RM} (for noise that does not have the symmetry 
of the maps) and for modeling complex quantum dynamics that have inherent
symmetries.

\begin{figure}
\includegraphics[height=5.8cm, width=8cm]{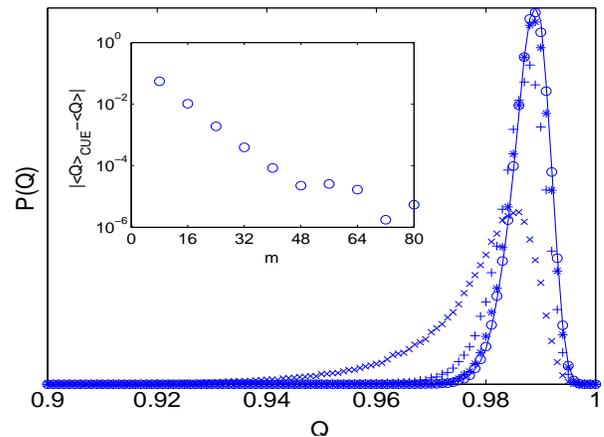}
\caption{\label{QQCA2} 
Distribution of the multi-partite entanglement, $Q$, for two-species QCA maps 
of $m = 16$ (x), 24, (+), 32 (*), and 40~(o) compared to the distribution for 
CUE random operators. As in the case of a one species QCA, the $m = 40$ 
distribution is barely distinguishable from the CUE distribution, 
despite the constraints imposed by the QCA architecture. The inset shows 
the difference between the average $Q$ for states from the QCA and 
CUE operators as a function of $m$. 
}
\end{figure}

\begin{figure}
\includegraphics[height=5.8cm, width=8cm]{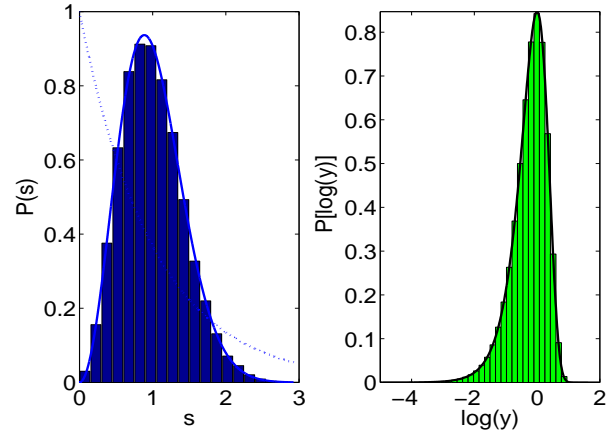}
\caption{\label{MQCA2} 
Statistical measures of randomness for 100 two-species, $m = 40$,
QCA maps. The left figure shows the distribution of $s$ the nearest neighbor 
eigenangle spacings compared to that expected for CUE operators (solid line), 
and regular integrable maps (dotted line). The right figure 
shows the distribution of the magnitude of the 
eigenvector elements $y$ compared to that expected for CUE (solid line). 
Both criteria of randomness are fulfilled by the $k = 2$ map. 
}
\end{figure}

The failure of the above operators to fulfill certain criteria of randomness 
can be rectified by breaking the mirror symmetry of the system. We provide two
examples. The first is by having a two species QCA chain
$ABAB \dots$ with an even number of qubits. For this $k = 2$ map, 
iteration $i$ of the circuit consists of a
random SU(2) rotation $U^A_i$ on all qubits of species $A$, followed by a
random SU(2) rotation $U^B_i$ applied to all qubits of species $B$, followed
by coupling between nearest neighbor qubits, $U_{nnc}$. Thus, the number 
of independent variables for the generated operator is $6m$ for the 
two random rotations per iteration plus one more for the coupling. As 
shown in figures (\ref{QQCA2}) and (\ref{MQCA2}) all of the tested 
criteria of randomness are fulfilled for these maps.

A second way to break the mirror symmetry of the $k = 1$ QCA is by changing 
the value of one of the nearest neighbor couplings (though not the center 
coupling). In this way the map requires only $3m+2$ independent variables 
while still fulfilling all of the above criteria of randomness. 
An operator generated from a system with unequal nearest neighbor 
couplings is especially relevant for experimental implementations. For 
many quantum computer hardware proposals attaining exact equal 
couplings between qubits is nearly impossible. An actual chain 
of qubits, such as in nuclear magnetic resonance or quantum dots, 
could not be expected to have the mirror symmetry assumed above.  
As we have demonstrated, however, this allows such systems to more 
easily generate pseudo-random states and operators. 

\begin{figure}
\includegraphics[height=5.8cm, width=8cm]{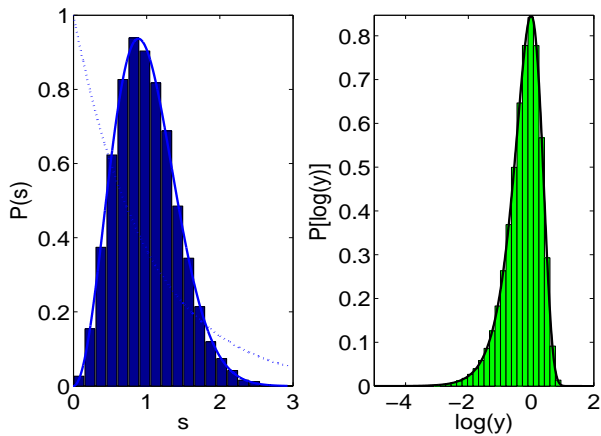}
\caption{\label{RingAsy2} 
Statistical measures of randomness for 100 $m = 40$, one-species QCA rings.
In this instance, all nearest neighbor couplings are equal to $\pi/4$ 
except for two, one of which has a coupling of $\pi/5$ and the other a 
coupling of $\pi/4.5$. The left figure shows the 
distribution of $s$ the nearest neighbor eigenangle spacings compared 
to that expected for CUE operators (solid line) and regular integrable maps 
(dotted line). The right figure shows the distribution of the magnitude 
of the eigenvector elements $y$ compared to the CUE distribution. 
QCA architecture with unequal nearest neighbor couplings 
(be it a ring or chain) are the most likely for experimental implementations.
}
\end{figure}

The discussion to this point has centered around QCA chains. If the QCA 
were structured as a ring (which, experimentally, may be more difficult) 
such that all qubits had two nearest neighbors, we have checked that 
the symmetries inherent in the system do not allow for enough entanglement 
production following the random distribution of $Q$, regardless of the 
number of iterations. However, if one of the couplings 
is (even slightly) different than the others the random distribution of $Q$ 
is recovered as the system is then equivalent to the QCA chain. If two 
couplings are different (from the others and each other), all symmetries 
have been broken and full pseudo-randomness is recovered as shown in figure 
(\ref{RingAsy2}).

There are other ways to reduce the number of independent variables in the 
pseudo-random operator algorithm besides a QCA. We 
explore some of these possibilities in attempt to achieve aspects 
of randomness with as few independent variables as possible. First, 
we explore what is, in some sense, the opposite of the QCA discussed 
above. For the QCA operators the same rotation was applied to every 
qubit but the rotation was different for each iteration. We now study 
operators in which a different random rotation is applied to each qubit
but the rotation is the same at every iteration. As before there is a 
$\pi/4$ nearest neighbor coupling evolution between 
the rotations. These operators, to which we shall refer to as repeat maps, 
require only $3n+1$ independent variables. 

Given the structure of repeat maps, the repetition of the same operation over 
and over, it is clear that they have time-reversal invariance, applying 
the map backwards produces the same results as applying it forward. Hence, 
the eigenvalue and eigenvector statistics deviate only slightly from 
the distribution appropriate for random orthogonal matrices. The circular 
orthogonal ensemble (COE) includes unitary operators that have anti-unitary 
symmetry (time reversal invariance) and is a subset of the 
general CUE. The level spacing for the COE class is \cite{Mehta}:
\begin{equation}
P_{COE}(s) = \frac{\pi s}{2}e^{\pi s^2/4},
\end{equation}
and the distribution of the elements of the eigenvectors of COE matrices 
as $N\rightarrow\infty$ are \cite{Mehta}
\begin{equation}
P_{COE}(y) = \frac{1}{\sqrt{2\pi y}}e^{-y/2}.
\end{equation} 
The operators of many quantum analogs of classically 
chaotic systems belong to the COE class and, therefore, repeat maps 
may form appropriate models for these systems. However, repeat maps do not 
produce the entanglement distribution expected for random operators.

\begin{figure}
\includegraphics[height=5.8cm, width=8cm]{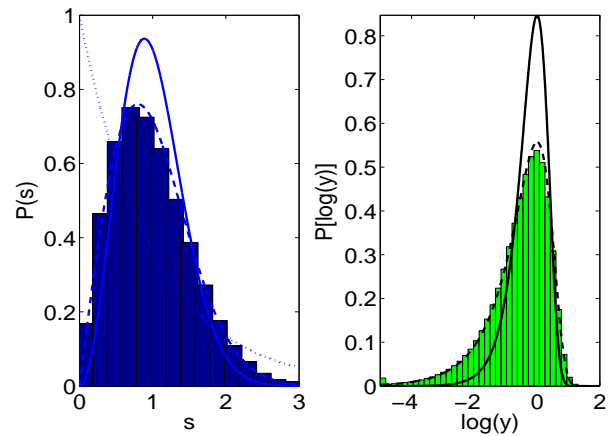}
\caption{\label{MRepeat} 
Statistical measures of randomness for 100 $m = 40$, repeat maps.
The left figure shows the distribution of 
$s$ the nearest neighbor eigenangle spacings compared to the expected 
distributions for CUE operators (solid line), regular integrable systems (
dotted line) and COE operators (dashed line). The right figure shows the 
distribution of the magnitude of the eigenvector elements $y$ compared
to the CUE (solid line) and COE (dashed line) distributions. Both 
distributions are similar to those of the COE universal symmetry class.
}
\end{figure}

Finally, we turn to versions of the pseudo-random operator algorithm which
faithfully follow the evolution of CCA. As explained, there are two major
characteristics of CCA evolution: homogeneity of evolution for each cell,
and homogeneity of evolution at each time step. In the first part of this
work, we examined evolution in accordance with only the first of these
characteristics. Repeat maps describe evolution with only the 
second characteristic. Currently, we explore the dynamics
of maps that evolve as CCA in both respects: each qubit rotates 
via the same random SU(2) rotation, $U$, and that rotation 
is the same for all iterations. This gives a total of only $3+1$ 
independent variable for the entire operator. 

\begin{figure}
\includegraphics[height=5.8cm, width=8cm]{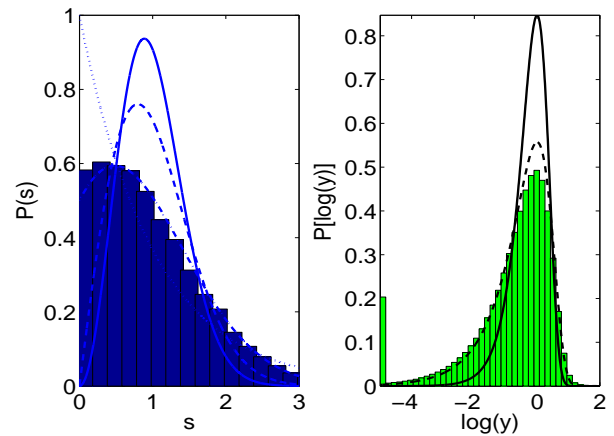}
\caption{\label{MQCARepeat} 
Statistical measures of randomness for 100 $m = 40$, $k = 1$, QCA maps in which
there is only one random rotation. This one rotation is applied to each qubit 
at each iteration. The left figure shows the distribution of 
$s$ the nearest neighbor eigenangle spacings compared to the global spectral
statistics of a COE operator with mirror symmetry (dash-dot line), 
and the right figure shows the distribution of the magnitude of the 
eigenvector elements $y$. Both distributions are similar to those of the 
COE universal symmetry class.
}
\end{figure}

As these QCA maps are even more limited than the repeat maps discussed 
above, it is no surprise that the entanglement produced by these maps do 
not follow the distribution of random maps. However, these maps deviate 
only slightly in the other criteria of randomness. As shown in 
figure (\ref{MQCARepeat}) the eigenvector element distribution deviates 
somewhat from the COE distribution while the nearest-neighbor spacing 
distribution deviates only slightly from the global statistics expected
from a map with two differently sized COE blocks due to mirror symmetry 
\cite{spec2}
\begin{eqnarray}
P_{COE}^{(2)}(s,g_1,g_2) &=& \frac{\pi}{2} sg_1^3{\rm
erfc}\left(\frac{\sqrt{\pi}g_2s}{2}\right)e^{-\pi g_1^2s^2/4}\nonumber\\
&+& \frac{\pi}{2} sg_2^3{\rm erfc}\left(\frac{\sqrt{\pi}g_1s}{2}\right)e^{-\pi
g_2^2s^2/4}\nonumber\\
&+& 2 g_1g_2e^{-\pi s^2(g_1^2+g_2^2)/4},
\end{eqnarray}
with $g_1$ and $g_2$ defined as above. As with the CUE mirror symmetry 
operator, the resulting distribution is barely distinguishable from the 
$g_1 = g_2 = 1/2$ case
\begin{equation}
P_{COE}^{(2)}(s) = \frac{1}{2}\big({\rm erfc}\left(\frac{\sqrt{\pi}s}{4}\right)\frac{\pi s}{4}e^{-\pi s^2/16}+e^{-\pi s^2/8}\big)
\end{equation}

\begin{figure}
\includegraphics[height=5.8cm, width=8cm]{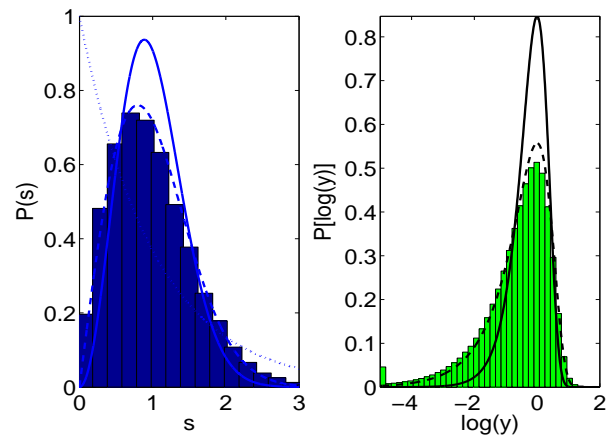}
\caption{\label{MQCARepeatAsy} 
Statistical measures of randomness for 100 $m = 40$, $k = 1$, QCA maps in 
which the same random rotation is applied to each qubit at each iteration. 
In these maps one of the nearest neighbor couplings in the chain
is $\pi/5$ while all the rest are $\pi/4$. The left figure shows 
the distribution of $s$ the nearest neighbor eigenangle spacings 
compared to COE, and the right figure shows the distribution of the 
magnitude of the eigenvector elements $y$ compared to that expected to 
COE. 
}
\end{figure}

As with the QCA maps discussed at the beginning of this work, the mirror
symmetry of the system can be broken by changing one of the couplings. In 
this way the eigenvalue and eigenvector statistics revert back to the 
COE distributions seen in the repeat map and shown in figure 
(\ref{MQCARepeatAsy}). This operator requires only $3+2$ independent variables.

Perhaps it should come as no surprise that operators with so few independent 
variables can still fulfill criteria of randomness. Quantum chaos models, 
such as the quantum sawtooth and quantum Harper's map \cite{qkt,Harper}, 
have only one or two free parameters, yet, fulfill criteria of randomness, and 
can be efficiently implemented on a quantum computer. What we 
have shown here, however, is that quantum chaos models are not exceptional 
cases, showing randomness due to their connection with a classically
chaotic analog. Rather, most operators with few random variables will 
still show many characteristics of randomness. Regularity is the exception,
randomness is the general rule.

In conclusion, we have demonstrated that general aspects of complex 
quantum dynamics can be studied on a QCA architecture. The evidence behind 
this supposition is the ability to  generate pseudo-random operators that 
fulfill general criteria of randomness. These operators are appropriate 
substitutes for random operators which are often used as models of complex 
quantum systems and play an important role in other quantum information 
processing protocols. Throughout this work we have attempted to introduce 
operators generated by as few independent variables as possible. These
operators are tested to determine whether they meet the  various criteria 
of randomness. Operators which do not fulfill all of the criteria may 
nevertheless prove useful in certain simulations of quantum systems and
other quantum computational protocols. The minimum number of independent 
variables is reached via simulation of an algorithm which mimics classical 
cellular automata evolution. Yet, even these operators demonstrate many 
aspects of randomness. This implies that even with few independent variables 
most operators will tend towards randomness.

The authors would like to thank S. Montangero for helpful discussions.
The authors acknowledge support from the DARPA QuIST (MIPR 02 N699-00)
program. Y.S.W. acknowledges the support of the National Research Council 
Research Associateship Program through the Naval Research Laboratory.
Computations were performed at the ASC DoD Major Shared Resource Center.

\end{document}